\documentclass[aps,twocolumn,showpacs,pre]{revtex4-1}
\usepackage{graphicx,graphics}
\usepackage{amsmath,amssymb,amsfonts}
\usepackage{bbm,bbold}
\usepackage{multirow}
\usepackage{hyperref}
\usepackage{graphicx}
\usepackage{mathptmx}
\usepackage{url}

\begin{document}
\title{\Large Mean encounter times for multiple random walkers on networks}
\author{Alejandro P.~Riascos}
\email{aperezr@fisica.unam.mx}
\affiliation{Instituto de F\'isica, Universidad Nacional Aut\'onoma de M\'exico, Ciudad Universitaria, Ciudad de M\'exico 04510, Mexico}
\author{David P.~Sanders} 
\email{dpsanders@ciencias.unam.mx}
\affiliation{Departamento de F\'isica, Facultad de Ciencias, Universidad Nacional Aut\'onoma de M\'exico, Ciudad Universitaria,  Ciudad de M\'exico 04510, Mexico \\ \& Department of Mathematics, Massachusetts Institute of Technology, Cambridge MA 02139, USA}
\date{\today}
\begin{abstract}
We introduce a general approach for the study of the collective dynamics of non-interacting random walkers on connected networks. We analyze the movement of $R$ independent (Markovian) walkers, each defined by its own transition matrix. By using the eigenvalues and eigenvectors of the $R$ independent transition matrices, we deduce analytical expressions for the collective stationary distribution and the average number of steps needed by the random walkers to start in a particular configuration and reach  specific nodes the first time (mean first-passage times), as well as global times that characterize the global activity. We apply these results to the study of mean first-encounter times for local and non-local random walk strategies on different types of networks, with both synchronous and asynchronous motion.
\end{abstract}

\maketitle
\section{Introduction}
The study and understanding of dynamical processes taking place on networks have had a significant impact with important contributions in science \cite{VespiBook,barabasi2016book,NewmanBook}. In particular, the dynamics of a random walker that visits the nodes of networks following different strategies is a challenging theoretical problem where the relation between network topology and the way the walker hops between nodes is explored \cite{Hughes,MasudaPhysRep2017,riascos2019random,GiuggioliPRX202}. Local strategies, where random walkers move from a node to one of its nearest neighbors, include normal random walks \cite{NohRieger} and degree biased random walks \cite{FronczakPRE2009}, among others \cite{BurdaPRL2009,SinatraPRE2011,LambiottePRE2011,Zhang2011,ZhangPRE2013}. In contrast, non-local random walks use global information of the network structure with a dynamics that allows long-range transitions, like the Google random walker \cite{Brin1998}, L\'evy flights on networks \cite{RiascosMateos2012,Weng2015,Guo2016,Estrada2017Multihopper}, fractional diffusion \cite{RiascosMateosFD2014,deNigris2017,RiascosMichelitsch2017_gL,FractionalBook2019,Allen_Perkins_2019}, and random walks with reset \cite{ResetRWnet2020}. Different developments in the understanding of random walkers on networks have led to valuable tools in searching processes on the internet \cite{Brin1998,ShepelyanskyRevModPhys2015}, algorithms for data mining \cite{BlanchardBook2011, LeskovecBook2014}, the understanding of human mobility in cities \cite{Barbosa2018,RiascosMateosPlosOne2017,LoaizaMonsalvePlosOne2019, RiascosMateosSciRep2020}, epidemic spreading \cite{BrockmannPRX2011,ValdezBraunsteinHavlin2020}, algorithms for image analysis \cite{GradyIEEE2006,GradyIEEE2007}, and unsupervised classification algorithms \cite{SdeNigris2017,Bautista2019}, just to mention a few applications.
\\[2mm]
\begin{figure}[!b] 
	\begin{center}
		\includegraphics*[width=0.475\textwidth]{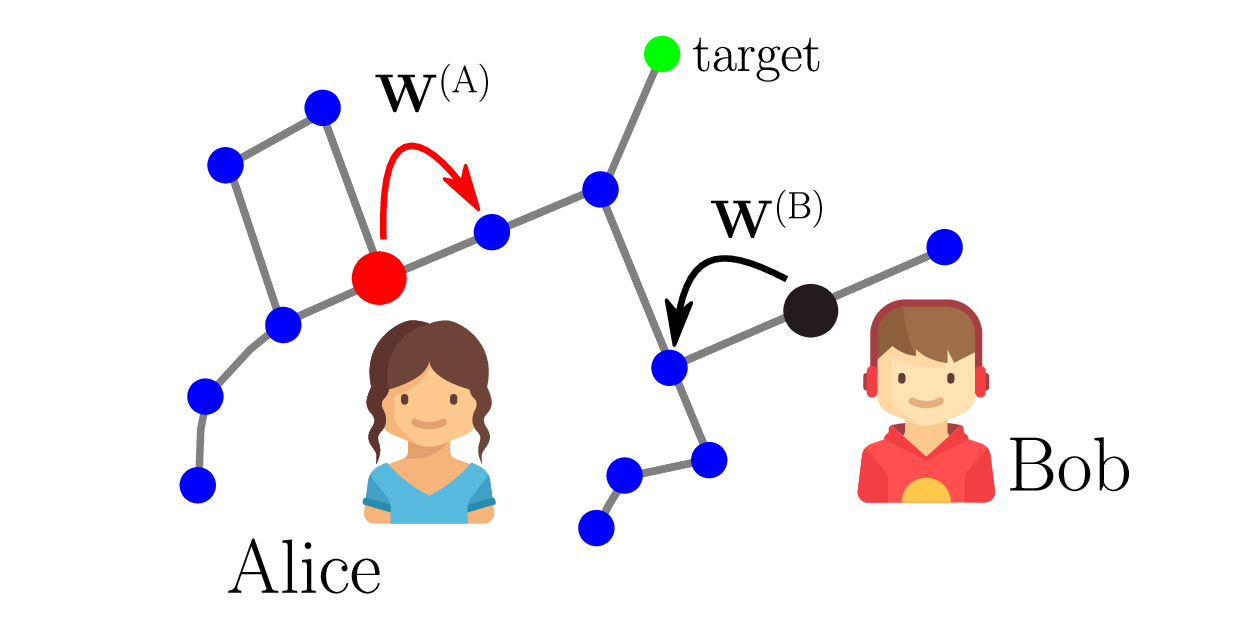}
	\end{center}
	\vspace{-8mm}
	\caption{\label{Fig_1}(Color online) Two random walkers visiting places represented by nodes in a street network. Alice visits sites by hopping between nodes with probabilities defined by a transition matrix $\mathbf{W}^{(\mathrm{A})}$, whereas Bob visits locations with a different strategy defined by $\mathbf{W}^{(\mathrm{B})}$. } 
\end{figure}
Most of the above-mentioned studies explore the dynamics of a single random walker; the dynamics of multiple walkers moving simultaneously have been less extensively considered \cite{WengPRE2017}. Multiple walkers (agents) are commonly found in real processes on complex systems; for example in encounter networks in human activity \cite{RiascosMateosPlosOne2017,Mastrandrea_PlosOne2015}, epidemic spreading \cite{SatorrasPRL2001,ValdezBraunsteinHavlin2020}, ecology \cite{GiuggioliPRL2013,CraftRoyalB_2015}, extreme events \cite{KishorePRL2011}, among others. Despite these potential applications, a complete theoretical framework for the analysis of simultaneous random walkers is still missing. Some of the recent advances consider the searching efficiency of multiple walkers on networks, exploring the mean time required to find a given target by one or some of the walkers \cite{DaiPhysA2020}, universal laws governing the search time  \cite{WengPRE2017, WengPRE2018universal, DaiPhysA2020}, analytical results for encounter times for many random walkers \cite{DPSandersPRE2009} and the expected time searchers take to capture moving targets
specified in advance \cite{Weng_2017, WengChaos2018}. Figure \ref{Fig_1} illustrates some of the possible situations that arise when we consider the activity of two agents visiting nodes following edges represented by lines. Even if the two walkers never interact with one another, it is important to know if these walkers coincide, or \emph{encounter} one another, the average time to reach for the first time particular nodes or to meet at a particular target node. All these cases are highly influenced by the network structure, the initial conditions, and how each random walker movement is defined. A theoretical understanding of the collective dynamics of simultaneous random walkers would have applications in human mobility and urban planning, epidemic spreading, ecology, among others.
\\[2mm]
In this work, we develop a general framework to study the collective movement of $R$ synchronous and asynchronous non-interacting random walkers, each defined by its own transition matrix, finding general, exact expressions describing the global activity of the random walkers. We analyze the stationary distribution and the average time to reach a particular set of nodes from given initial conditions. The analytical results are expressed in terms of the eigenvalues and eigenvectors of the individual transition matrices defining each random walker. We explore results for two walkers following local (normal, degree biased) and non-local (L\'evy flight) random walk strategies on different networks, including trees, combs, rings, and random networks. We also explore the effect of the initial conditions on mean first-encounter times for path and ring graphs for up to five walkers. Finally, the results are applied to study the activity of the bike-sharing system Citibike in New York City, where we explore encounter times of two and three bikes at each station; this example illustrates possible applications of our formalism in the context of transportation systems and human mobility.
\section{General theory}
\label{Sec_General}
\subsection{Master equation}
We study the activity of $R$ random walkers on a general connected network (graph) with $N$ nodes, $\mathcal{V}=\{1,2,\ldots,N\}$, given by an adjacency matrix $\mathbf{A}$ with elements $A_{ij}$, i.e., such that nodes $i$ and $j$ are joined by an edge if and only if $A_{ij} = 1$. Walker $r$ is defined by an $N \times N$ transition matrix $\mathbf{W}^{(r)}$, where the element  $(\mathbf{W}^{(r)})_{ij}=w^{(r)}_{i\to j}$ determines the probability to hop from node $i$ to $j$ (with $r=1, 2, \ldots, R$).  At discrete times $t = 1, 2, \ldots$, walkers hop independently. We study two possible dynamics: \emph{synchronous}, where all random walkers jump simultaneously, and \emph{asynchronous}, where a single one of the $R$ walkers is chosen at random to move (i.e., each walker is chosen with equal probability $1/R$).
\\[2mm]
A matrix ${\mathcal{W}}$ that describes the global activity of these $R$ non-interacting random walkers is given, for synchronous motion, by
\begin{equation}\label{Wmatrix_S}
{\mathcal{W}^\mathrm{S}} \equiv \bigotimes_{r=1}^R \mathbf{W}^{(r)}
= \mathbf{W}^{(1)} \otimes \mathbf{W}^{(2)} \otimes \cdots \otimes \mathbf{W}^{(R)},
\end{equation}
where $\otimes$ denotes the tensor product (Kronecker product) of matrices.
For asynchronous motion, in which a single walker moves at each time step, the dynamics is instead given by
\begin{multline}
{\mathcal{W}^\mathrm{A}}\equiv 
\frac{1}{R} \Big[ \mathbf{W}^{(1)} \otimes \mathbf{I} \otimes \cdots  \otimes \mathbf{I} \, + \, \mathbf{I} \otimes \mathbf{W}^{(2)} \otimes \mathbf{I} \otimes \cdots \otimes \mathbf{I} \\
 + \cdots + \, \mathbf{I} \otimes \cdots
\otimes \mathbf{I} \otimes \mathbf{W}^{(R)} 
\Big].\label{Wmatrix_A}
\end{multline}
It is convenient to introduce the notation $ \vec{i}\equiv(i_1,i_2,\ldots,i_R)\in \mathcal{V}^R$, with $i_1, i_2, \ldots, i_R = 1,2,\ldots, N$, for a vector describing the positions of each  walker, where $i_r$ is the position (node) of  walker $r$ on the network. The probability $\mathcal{P}(\vec{i},\vec{j};t)$ to find the $R$ synchronous walkers respectively at nodes $\vec{j}$ at time $t$, starting from initial positions $\vec{i}$ at $t=0$, is then given by
\begin{equation}\label{PsetRW}
\mathcal{P}(\vec{i},\vec{j};t) \equiv P^{(1)}_{i_1 j_1}(t) \, P^{(2)}_{i_2 j_2}(t) \, \cdots \, P^{(R)}_{i_R j_R}(t),
\end{equation}
where $P^{(r)}_{ij}(t)$ is the occupation probability to find the $r$th  walker at the node $j$ at time $t$, starting from $i$ at $t=0$. By definition, each of these individual occupation probabilities satisfies the master equation \cite{Hughes,NohRieger}
\begin{equation}\label{master_singleRW}
P^{(r)}_{i j}(t+1) = \sum_{m=1}^N  P^{(r)}_{i m} (t) w^{(r)}_{m\rightarrow j}, \qquad r=1,2,\ldots, R.
\end{equation}
Using the canonical basis of $\mathbb{R}^N$, written in Dirac notation as $\{\left| i\right\rangle\}_{i=1}^N$, Eq.~(\ref{master_singleRW}) leads to $P^{(r)}_{ij}(t)=\left\langle i\right| (\mathbf{W}^{(r)})^t\left|j\right\rangle$. Therefore, we have for the probability $\mathcal{P}(\vec{i} ,\vec{j};t)$ in Eq.~(\ref{PsetRW})
\begin{equation}\label{PMnoninterS}
\mathcal{P}(\vec{i},\vec{j};t)=\prod_{r=1}^R \langle i_r|(\mathbf{W}^{(r)})^t|j_r\rangle=\langle \vec{i}|{(\mathcal{W}^\mathrm{S})}^t|\vec{j}\rangle,
\end{equation}
where we use the compact notation
\begin{equation}
|\vec{i}\rangle\equiv
|i_1,i_2,\ldots,i_R\rangle=|i_1\rangle\otimes|i_2\rangle\otimes \cdots \otimes|i_R\rangle.
\end{equation}
The matrix ${\mathcal{W}^\mathrm{S}}$ describes the collective movement of $R$ synchronous non-interacting random walkers; in this way, the elements $\mathcal{W}_{\vec{i} \; \vec{j}}\equiv\langle \vec{i}|{\mathcal{W}}|\vec{j}\rangle$ define the transition probability between the  configuration described by the vector $\vec{i}$ to a new state $\vec{j}$. These transitions have the structure of a stochastic matrix for a Markovian process, where the possible configurations are now given by vectors in $\mathcal{V}^R$, denoting the positions of all $R$  walkers on the network. Since the individual transition matrices $\mathbf{W}^{(r)}$ are stochastic, ${\mathcal{W}^\mathrm{S}}$ satisfies  
\begin{equation}\label{stoch_tensor}
\sum_{\vec{j}\in \mathcal{V}^R}\mathcal{W}^\mathrm{S}_{  \vec{i}\; \vec{j}}
=\sum_{\vec{j}\in \mathcal{V}^R}\langle \vec{i}|{\mathcal{W}^\mathrm{S}}|\vec{j}\rangle
=\prod_{r=1}^R\sum_{j_r=1}^N w_{i_r\to j_r}^{(r)}=1.
\end{equation}
Similar relations can be found for the asynchronous motion defined by  Eq. (\ref{Wmatrix_A}). Then, the simultaneous dynamics satisfies
\begin{equation}\label{PMnoninter}
\mathcal{P}(\vec{i},\vec{j};t)=\langle \vec{i}|{\mathcal{W}}^t|\vec{j}\rangle,
\end{equation}
for $\mathcal{W}$ in Eqs.~(\ref{Wmatrix_S})--(\ref{Wmatrix_A}). Furthermore, from Eq.~(\ref{PMnoninter}), the probability $\mathcal{P}(\vec{i} ,\vec{j};t)$ evolves according to the master equation 
\begin{align}\nonumber
\mathcal{P}(\vec{i} ,\vec{j};t+1)&=\langle \vec{i}|{\mathcal{W}}^{t+1}|\vec{j}\rangle=\sum_{\vec{l}\in \mathcal{V}^R} \langle \vec{i}|{\mathcal{W}}^{t}|\vec{l}\rangle \langle \vec{l}|{\mathcal{W}}|\vec{j}\rangle\\
&=\sum_{\vec{l}\in \mathcal{V}^R} \mathcal{P}(\vec{i} ,\vec{l};t) \mathcal{W}_{\vec{l} \; \vec{j}} \; .
\label{masterRrw}
\end{align}
An equivalent alternative viewpoint is to regard the movement of the $R$ walkers as a single walker on a particular product graph \cite{Patel2016SIAM}.

\subsection{Spectral form and stationary distribution}
Equations (\ref{stoch_tensor})--(\ref{masterRrw}) have an exact parallel with the dynamics of a single walker. In this way,  it is possible to treat the problem of $R$ non-interacting random walkers analytically; in particular, to calculate the mean number of steps needed to reach a given configuration. To do so, let us firstly introduce a compact notation for the eigenvalues and eigenvectors of ${\mathcal{W}}$, since these are key quantities in the study of the master equation (\ref{masterRrw}). They can be deduced directly from the analysis of the matrices $\mathbf{W}^{(r)}$. We will suppose that each individual transition matrix $\mathbf{W}^{(r)}$ is diagonalizable. A sufficient condition for this is that each satisfies detailed balance, $P^{r,\infty}_{i}\, w^{(r)}_{i \rightarrow j} = P^{r,\infty}_{j}\, w^{(r)}_{j \rightarrow i}$ with respect to its stationary distribution $P^{r,\infty}_{j}$. We then have %
\begin{equation}\label{Weigenvector}
\mathbf{W} ^{(r)}|\phi_i^{(r)}\rangle=\lambda_i^{(r)}|\phi_i^{(r)}\rangle \qquad r=1,2, \ldots, R,
\end{equation}
where $\{\lambda_i^{(r)}\}_{i=1}^{N}$ denote the eigenvalues of the transition matrix $\mathbf{W} ^{(r)}$ with the corresponding set of right eigenvectors $\{ |\phi_i^{(r)}\rangle\}_{i=1}^{N}$ \cite{riascos2019random}. In terms of these eigenvectors we define 
\begin{equation}
|\phi_{\vec{i}}\rangle\equiv|\phi_{i_1}^{(1)}\rangle\otimes|\phi_{i_2}^{(2)}\rangle\otimes \cdots \otimes|\phi_{i_R}^{(R)}\rangle=\bigotimes_{r=1}^R |\phi_{i_r}^{(r)}\rangle
\end{equation}
and, combining this definition with Eqs.~(\ref{Wmatrix_S}), (\ref{Wmatrix_A}) and (\ref{Weigenvector}), we obtain
\begin{equation}\label{SpectM}
{\mathcal{W}}|\phi_{\vec{i}}\rangle=\zeta_{\vec{i}}|\phi_{\vec{i}}\rangle,
\end{equation}
where, using the definition in Eq.~(\ref{Wmatrix_S}) for synchronous random walkers, we obtain eigenvalues
\begin{equation}\label{zeta_S}
\zeta^\mathrm{S}_{\vec{i}} \equiv \prod_{r=1}^R \lambda_{i_r}^{(r)},
\end{equation}
 of ${\mathcal{W}^\mathrm{S}}$. Similarly, for the asynchronous motion with ${\mathcal{W}^\mathrm{A}}$ defined by Eq.~(\ref{Wmatrix_A}) we obtain
\begin{equation}\label{zeta_A}
\zeta^\mathrm{A}_{\vec{i}} \equiv \frac{1}{R}\sum_{r=1}^R \lambda_{i_r}^{(r)},
\end{equation}
using the fact that the individual operators in Eq.~(\ref{Wmatrix_A}) commute and thus are simultaneously diagonalizable.
In addition, the eigenvalues of $\mathbf{W}^{(r)}$ satisfy $|\lambda_{i_r}^{(r)}|\leq 1$; therefore $|\zeta_{\vec{i}}| \leq 1$ for all ${\vec{i}}\in \mathcal{V}^R$ for both cases in Eqs.~(\ref{zeta_S}) and (\ref{zeta_A}).
\\[2mm]
We will also require left eigenvectors. For the  individual transition matrix $\mathbf{W}^{(r)}$ we have $ \langle \bar{\phi}^{(r)}_i|\mathbf{W}^{(r)}= \lambda_i^{(r)}\langle \bar{\phi}^{(r)}_i|$; for the full dynamics  $\langle\bar{\phi}_{\vec{i}}| \equiv \bigotimes_{r=1}^R \langle\bar{\phi}_{i_r}^{(r)}|$ satisfies
\begin{equation}
\langle\bar{\phi}_{\vec{i}}|{\mathcal{W}}=
\zeta_{\vec{i}}\langle\bar{\phi}_{\vec{i}}|.
\end{equation}
Each set of eigenvectors of $\mathbf{W}^{(r)}$ satisfies $\delta_{ij}=\langle\bar{\phi}^{(r)}_i|\phi^{(r)}_j\rangle$ and $\mathbb{1}=\sum_{l=1}^N |\phi^{(r)}_l\rangle \langle \bar{\phi}^{(r)}_l |$, where $\delta_{ij}$ is the Kronecker delta and $\mathbb{1}$ is the $N\times N$ identity matrix \cite{riascos2019random,FractionalBook2019}. Therefore, from the definitions of $|\phi_{\vec{i}}\rangle$ and $\langle\bar{\phi}_{\vec{j}}|$, we have the orthonormalization condition
\begin{equation}
\langle\bar{\phi}_{\vec{i}}|\phi_{\vec{j}}\rangle=\delta_{i_1,j_1}\delta_{i_2,j_2}\ldots\delta_{i_R,j_R}
\equiv \delta_{\vec{i},\vec{j}}
\end{equation}
and the completeness relation $\sum_{\vec{l}\in \mathcal{V}^R}\left|\phi_{\vec{l}}\right\rangle\left\langle\bar{\phi}_{\vec{l}}\,\right|=\mathbb{1}^{\otimes R}$. 
\\[2mm]
In the following,  we denote the maximum eigenvalue of $\mathbf{W}^{(r)}$ as $\lambda_1^{(r)}=1$; this eigenvalue is unique according to the Perron--Frobenius theorem and the corresponding eigenvector defines the stationary distribution of each  walker through the relation $P^{r,\infty}_{j}=\langle i|\phi_1^{(r)}\rangle \langle\bar{\phi}^{(r)}_1|j\rangle$,  independent of the initial node $i$ since $\langle i|\phi_1^{(r)}\rangle$ is a constant \cite{riascos2019random}.
\\[2mm]
We can now express the time evolution $\mathcal{P}(\vec{i} ,\vec{j};t)$ of the $R$-walker system in terms of the eigenvalues and left and right eigenvectors of ${\mathcal{W}}$, as follows. From Eq.~(\ref{PMnoninter}) we have
\begin{align}\nonumber
\mathcal{P}(\vec{i} ,\vec{j};t)&=\langle \vec{i}|{\mathcal{W}}^t|\vec{j}\rangle
= \sum_{\vec{l}\in \mathcal{V}^R}\langle \vec{i}|{\mathcal{W}}^t\left|\phi_{\vec{l}}\right\rangle\left\langle\bar{\phi}_{\vec{l}}\,\right|\vec{j}\rangle\\ \label{Pijsimul_spect}
&=\sum_{\vec{l}\in \mathcal{V}^R}  \zeta_{\vec{l}}^t\langle \vec{i} |\phi_{\vec{l}}\rangle\langle\bar{\phi}_{\vec{l}}|\vec{j}\rangle.
\end{align}
Hence we obtain for the stationary distribution   $\mathcal{P}^{\infty}_{\vec{j}}(\vec{i})\equiv \lim_{T\to \infty}\frac{1}{T}\sum_{t=0}^{T} \mathcal{P}(\vec{i} ,\vec{j};t)$ 
\begin{align}\nonumber
\mathcal{P}^{\infty}_{\vec{j}}(\vec{i})&=\lim_{T\to \infty}\frac{1}{T}\sum_{t=0}^{T} \sum_{\vec{l}\in \mathcal{V}^R}  \zeta_{\vec{l}}^t\langle \vec{i} |\phi_{\vec{l}}\rangle\langle\bar{\phi}_{\vec{l}}|\vec{j}\rangle\\
&=\sum_{\vec{l}\in \mathcal{V}^R} \delta_{\zeta_{\vec{l}},1}\langle \vec{i} |\phi_{\vec{l}}\rangle\langle\bar{\phi}_{\vec{l}}|\vec{j}\rangle \, .	\label{Pinf_multiple}
\end{align}
In Eq.~(\ref{Pinf_multiple}) it is important to define the degeneracy of the eigenvalue $\zeta=1$. Considering the definition of the eigenvalues $\zeta_{\vec{i}}$ in Eqs.~(\ref{zeta_S})--(\ref{zeta_A}), we see that for synchronous random walkers it is possible that multiple eigenvectors could have the maximum eigenvalue $\max_{{\vec{i}}\in \mathcal{V}^R}\{\zeta^\mathrm{S}_{\vec{i}}\}=1$. In contrast, for the asynchronous case the maximum value of $\zeta^\mathrm{A}_{\vec{i}}$ is 1 and is unique, a consequence of having only one eigenvalue $\lambda_1^{(r)}=1$ for $r=1,2,\ldots,R$.
\\[2mm]
In the following, we denote by $\kappa\equiv \sum_{\vec{l}\in \mathcal{V}^R} \delta_{\zeta_{\vec{l}},1}$ the degeneracy of the largest eigenvalue of ${\mathcal{W}}$. In the case where $\kappa=1$, all initial configurations can lead to any possible final states in a finite time, i.e. the system is irreducible. On the other hand, when $\kappa>1$ there are initial conditions that cannot reach specific final states; in these cases the stationary distribution is zero. In particular, for a single random walker $r$, degeneracy of the largest eigenvalue of the transition matrix $\mathbf{W}^{(r)}$ occurs only when the network has disconnected parts.  We denote the set $\mathcal{D} 
\equiv \{\vec{l}\in\mathcal{V}^{R}: \zeta_{\vec{l}}=1\}$, and the complement $\mathcal{D}^\mathrm{c} \equiv \mathcal{V}^R\setminus\mathcal{D}$. The stationary distribution in Eq.~(\ref{Pinf_multiple}) then takes the form
\begin{equation}
\mathcal{P}^{\infty}_{\vec{j}}(\vec{i})=\sum_{\vec{l}\in \mathcal{D}} \langle \vec{i} |\phi_{\vec{l}}\rangle\langle\bar{\phi}_{\vec{l}}|\vec{j}\rangle .
\end{equation}
In cases with $\kappa>1$, the stationary distribution depends on the initial configuration $\vec{i}$, whereas for $\kappa=1$ we have
\begin{equation} \label{Pstationary_kappa1}
\mathcal{P}^{\infty}_{\vec{j}}=
P^{1,\infty}_{j_1}P^{2,\infty}_{j_2}\cdots P^{R,\infty}_{j_R}\, ,
\end{equation}
independent of the initial condition.
\subsection{Mean first-passage time}
We now calculate the average time $\langle T(\vec{i};\vec{j})\rangle\equiv\langle T(i_1,i_2,\ldots,i_R;j_1,j_2,\ldots,j_R)\rangle $ needed by the walkers to reach simultaneously for the first time the nodes described by the vector $\vec{j}$ if at time $t=0$ the initial nodes are  $\vec{i}$. The mathematical formalism necessary to deduce analytically this quantity is analogous to that for the mean first-passage time of a single random walker (see Refs.~\cite{Hughes,RiascosMateos2012,riascos2019random} for details). We center our discussion on the analysis of a Markovian process defined by the transition probabilities of $R$ non-interacting walkers described by ${\mathcal{W}}$ in Eqs.~(\ref{Wmatrix_S})--(\ref{Wmatrix_A}), satisfying the master equation in Eq.~(\ref{masterRrw}). For this case, the occupation probability $\mathcal{P}(\vec{i},\vec{j};t)$ can be expressed as
\begin{equation}\label{EquF}
\mathcal{P}(\vec{i},\vec{j};t)  = \delta_{t0} \delta_{\vec{i},\vec{j}} + \sum_{t'=0}^t  F(\vec{i},\vec{j};t')
\mathcal{P}(\vec{j},\vec{j};t-t'),
\end{equation}
where $F(\vec{i},\vec{j};t)$ is the first-passage probability to start in the configuration $\vec{i}=(i_1,i_2,\ldots,i_R)$ and reach the configuration $\vec{j}=(j_1,j_2,\ldots,j_R)$ for the first time after $t$ steps. Taking the discrete Laplace transform $\tilde{f}(s) \equiv\sum_{t=0}^\infty e^{-st} f(t)$ of Eq.~(\ref{EquF}) we obtain  
\begin{equation}\label{LaplTransF}
\widetilde{F}(\vec{i},\vec{j};s) = \frac{\widetilde{\mathcal{P}}(\vec{i},\vec{j};s) - \delta_{\vec{i},\vec{j}}}{\widetilde{\mathcal{P}}(\vec{j},\vec{j};s)}.
\end{equation}
The mean first-passage time (MFPT) $\langle T(\vec{i};\vec{j}\,)\rangle$ is then obtained via a series expansion of  $\widetilde{F}(\vec{i},\vec{j};s)$ in powers of $s$
\begin{equation}
\widetilde{F}(\vec{i},\vec{j};s)=1-s\langle T(\vec{i};\vec{j}\,)\rangle + \cdots,
\end{equation}
and using the stationary distribution $\mathcal{P}_{\vec{j}}^\infty(\vec{i}\,)$ we define the moments 
\begin{equation}\label{Rij_n}
\mathcal{R}^{(n)}(\vec{i},\vec{j})\equiv \sum_{t=0}^{\infty} t^n \left [\mathcal{P}(\vec{i},\vec{j};t)-\mathcal{P}_{\vec{j}}^\infty(\vec{i}\,) \right].
\end{equation}
In this way, the expansion of $\widetilde{\mathcal{P}}(\vec{i},\vec{j};s)$ is
\begin{equation}
\widetilde{\mathcal{P}}(\vec{i},\vec{j};s) = \frac{\mathcal{P}_{\vec{j}}^\infty(\vec{i}\,)}{(1-e^{-s})}
+ \sum_{n=0}^\infty (-1)^n \mathcal{R}^{(n)}(\vec{i},\vec{j}) \frac{s^n}{n!} \ .
\end{equation}
Substituting this result into Eq.~(\ref{LaplTransF}) and performing an expansion of $\widetilde{F}(\vec{i},\vec{j};s)$, we find
\begin{equation}\label{Tij_R}
\langle T(\vec{i};\vec{j}\,)\rangle=\frac{1}{\mathcal{P}_{\vec{j}}^\infty(\vec{i}\,)}\left[\mathcal{R}^{(0)}(\vec{j},\vec{j})-\mathcal{R}^{(0)}(\vec{i},\vec{j})+\delta _{\vec{i},\vec{j}}\right].
\end{equation}
Here the term with $\delta_{\vec{i},\vec{j}}$ gives the mean return time $\langle T(\vec{i};\vec{i}\,)\rangle=1/\mathcal{P}_{\vec{i}}^\infty(\vec{i}\,)$ to start in the configuration $\vec{i}$ and return for the first time to this particular state (the Kac lemma).
\\[2mm]
Now we use the spectral representation of  $\mathcal{P}(\vec{i},\vec{j};t)$ in Eq.~(\ref{Pijsimul_spect}) and the stationary distribution $\mathcal{P}_{\vec{j}}^\infty(\vec{i}\,)$ in Eq.~(\ref{Pinf_multiple}) to calculate $\langle T(\vec{i};\vec{j}\,)\rangle$. From the definition of $\mathcal{R}^{(n)}(\vec{i},\vec{j})$ we have
\begin{align}\nonumber
\mathcal{R}^{(0)}(\vec{i},\vec{j})&=\sum_{t=0}^\infty \left[ \mathcal{P}(\vec{i},\vec{j};t)-\mathcal{P}_{\vec{j}}^\infty(\vec{i}\,) \right]\\
\nonumber
&=\sum_{t=0}^\infty \sum_{\vec{l}\in\mathcal{V}^{R}}\left[ \zeta_{\vec{l}}^t-\delta_{\zeta_{\vec{l}},1}\right] \langle \vec{i} |\phi_{\vec{l}}\rangle\langle\bar{\phi}_{\vec{l}}|\vec{j}\rangle.
\end{align}
In terms of the set $\mathcal{D}=\{\vec{l}\in\mathcal{V}^{R}: \zeta_{\vec{l}}=1\}$, and the respective complement $\mathcal{D}^\mathrm{c}$, we have
\begin{equation}
\mathcal{R}^{(0)}(\vec{i},\vec{j})=\sum_{\vec{l}\in \mathcal{D}^\mathrm{c}}\frac{1}{1-\zeta_{\vec{l}}}\langle \vec{i} |\phi_{\vec{l}}\rangle\langle\bar{\phi}_{\vec{l}}|\vec{j}\rangle.
\label{Rij0}
\end{equation}
Finally, the introduction  of this result into Eq.~(\ref{Tij_R}) gives for $\vec{i}\neq\vec{j}$
\begin{equation}\label{TijSpect}
\langle T(\vec{i};\vec{j}\,)\rangle
=\frac{1}{\mathcal{P}_{\vec{j}}^\infty(\vec{i}\,)}
\sum_{\vec{l}\in \mathcal{D}^\mathrm{c}}\frac{\langle \vec{j} |\phi_{\vec{l}}\rangle\langle\bar{\phi}_{\vec{l}}|\vec{j}\rangle-
	\langle \vec{i} |\phi_{\vec{l}}\rangle\langle\bar{\phi}_{\vec{l}}|\vec{j}\rangle}{1-\zeta_{\vec{l}}}
\end{equation}
and $\langle T(\vec{i};\vec{i}\,)\rangle=1/\mathcal{P}_{\vec{i}}^\infty(\vec{i}\,)$.
\\[2mm]
The approach described in this section applies for both synchronous and asynchronous random walkers, depending on the choice of the eigenvalues $\zeta_{\vec{l}}$. For synchronous motion we choose $\zeta^{\mathrm{S}}_{\vec{l}}$ in Eq.~(\ref{zeta_S}), whereas the choice $\zeta^{\mathrm{A}}_{\vec{l}}$ in Eq.~(\ref{zeta_A}) gives asynchronous motion. The corresponding eigenvectors are the same in both cases. 
\section{Mean first-encounter times for synchronous random walkers}
In this section we apply the above analytical results to study different characteristics of synchronous  random walkers described by $\mathcal{W}^\mathrm{S}$ in Eq.~(\ref{Wmatrix_S}). Using this formalism, we analyze the mean first-encounter time, defined as the time to start at nodes $\vec{i}$ and coincide for the first time at node $j$, by evaluating Eq.~(\ref{Tij_R}) for $j_1=j_2=\cdots=j_R=j$, for different numbers and types of walkers on various graph types.
\subsection{Two normal random walkers}
\begin{figure*}[!t] 
	\begin{center}
		\includegraphics*[width=0.9\textwidth]{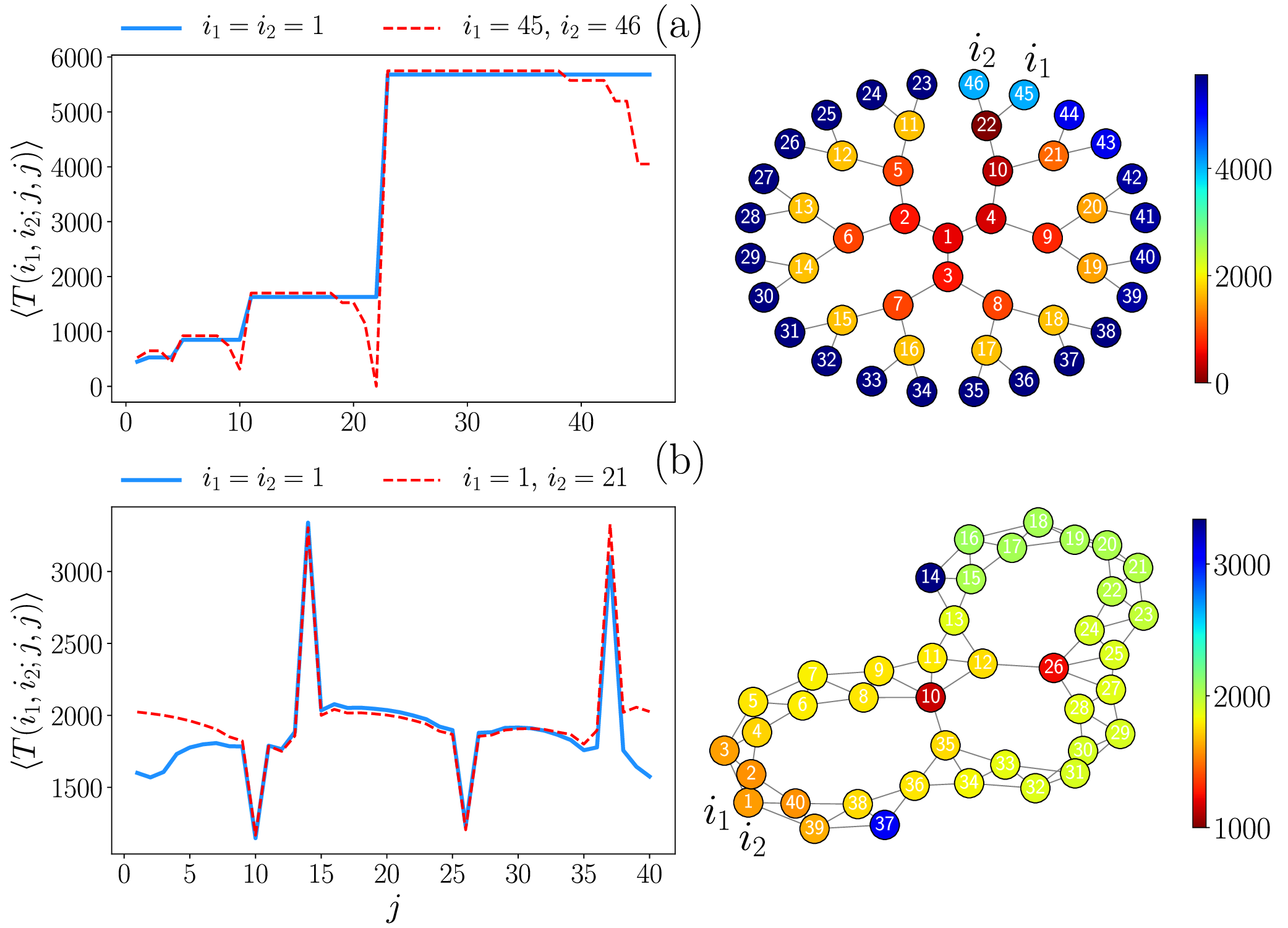}
	\end{center}
	\vspace{-8mm}
	\caption{\label{Fig_2}(Color online) Mean first-encounter times for two synchronous normal random walkers in connected networks: (a) a Cayley tree, (b) a Watts--Strogatz network. In the left panels, we show $\langle T(i_1,i_2;j,j)\rangle$ as a function of the node $j$ where the walkers coincide for the first time, for two different initial conditions. Each node $j$ of the networks is colored according to the mean time $\langle T(i_1,i_2;j,j)\rangle$ for the initial condition $i_1=45$, $i_2=46$ for the Cayley tree in (a) and $i_1=i_2=1$ for the Watts--Strogatz network in (b).} 
\end{figure*}
We proceed to apply Eqs.~(\ref{Pinf_multiple})--(\ref{TijSpect}) to calculate the mean time $\langle T(i_1,i_2;j,j)\rangle$ taken by $R=2$ standard (normal) random walkers that start at $t=0$ from nodes $i_1$ and $i_2$, respectively, to coincide for the first time at the node $j_1=j_2=j$. Each walker hops with an individual transition probability matrix $\mathbf{W}$, given in terms of the elements of the adjacency matrix $A_{lm}$ by $w_{l\to m} \equiv A_{lm}/k_l$, where $k_l \equiv \sum_m A_{lm}$ is the degree of node $l$; for this dynamics the (individual) stationary distribution is known to be $P_{j}^{\infty}=\frac{k_j}{\sum_{l=1}^N k_l}$ \cite{NohRieger}.  For this case, $\kappa=\sum_{l,m=1}^N \delta_{\lambda_l\lambda_m,1}$, hence we obtain $\kappa=2$ if the transition matrix has the eigenvalues $\lambda=\pm 1$. For normal random walks this occurs in bipartite networks \cite{VanMieghem2011,GodsilBook,FractionalBook2019}, a particular class of undirected graph having the property that the vertices can be partitioned into two disjoint sets with each link connecting only nodes in different sets; examples include cycles with an even number of nodes, and trees.  If the network is not bipartite then $\kappa=1$ (associated to $\lambda= 1$) and the stationary distribution in Eq.~(\ref{Pijsimul_spect}) is  $\mathcal{P}_{(j,j)}^{\infty}=(P_{j}^{\infty})^2$, independent of the initial node. Furthermore, 
Eq.~(\ref{Tij_R}) gives
\begin{multline}\label{MFETtworandomkappa1}
\langle T(i_1,i_2;j,j)\rangle =\frac{1}{(P_{j}^\infty)^2}\times
\Big[\delta_{i_1,j}\delta_{i_2,j}+\\\sum_{l,m=1}^N g(\lambda_l\lambda_m)\left(X_{jj}^{(l)}X_{jj}^{(m)}-X_{i_1j}^{(l)}X_{i_2j}^{(m)}\right)\Big]
\end{multline}
where $X_{ij}^{(l)}=\left\langle i|\phi_l\right\rangle \left\langle \bar{\phi}_l|j\right\rangle$ and 
\begin{equation}
g(z)\equiv\begin{cases}
(1-z)^{-1}, &\text{if } z\neq 1,\\
0,        &\text{if } z= 1.   
\end{cases}
\end{equation}
In Fig.~\ref{Fig_2} we show mean encounter times for two normal random walkers in a Cayley tree and in a random small-world network generated with the Watts--Strogatz algorithm \cite{WattsStrogatz1998}. In the left panels we present numerical results for $\langle T(i_1,i_2;j,j)\rangle$ for two different  initial conditions $(i_1, i_2)$, one in which the two walkers start from the same node and one in which $ i_1 \neq i_2 $. To illustrate the topology of the networks analyzed and the encounter times, in 
Fig.~\ref{Fig_2} we also show the networks, with nodes colored according to the encounter time for the initial condition $i_1=45$ and $i_2=46$ for the Cayley tree and $i_1=i_2=1$ for the Watts--Strogatz network.
\\[2mm]
In Fig.~\ref{Fig_2}(a) for a Cayley tree with $ N = 46 $ nodes we apply the  general equation in 
Eq.~(\ref{Tij_R}), since this is a bipartite network. The results reveal the differences between the two initial conditions; in particular, for $ i_1 = i_2 = 1 $ the average encounter times are the same for the nodes that are at the same distance from the central node. This symmetry in the encounter times changes for the initial conditions $ i_1 = 45 $ and $ i_2 = 46 $. Here, the biggest differences are seen in the encounter times for nodes along the same branch as $i_1$ and $i_2$ (i.e. $j = 4, 9,10,19-22,39-46$). In particular the first-encounter time at $ j = 22 $ is exactly one step.  On the other hand, the results in Fig.~\ref{Fig_2}(b) for the Watts-Strogatz network with $ N = 40 $ are calculated using 
Eq.~(\ref {MFETtworandomkappa1}), since in this case 
$\kappa = 1$. Our findings show the variations when we modify the initial conditions. However, in this network with the small-world property there are specific nodes that offer great connectivity to the entire structure, where the encounter times are shorter, e.g., $ j = 10$ and $j = 26$, and with little variations with the change of the initial conditions. The evaluation of the betweenness centrality, that gives high centralities to nodes that are on many shortest paths of other node pairs \cite{NewmanBook}, reveals that nodes $10$, $12$, $26$ and $35$ have the highest betweenness centrality. We also see that in this particular case, the encounter times are higher in nodes $ j = 14$ and $ j = 37$, nodes with low betweenness centrality. These results show how two synchronous random walkers coincide faster in nodes that can be reached from different routes on the network.
\subsection{L\'evy flights on networks}
\begin{figure*}[!t] 
	\begin{center}
		\includegraphics*[width=0.9\textwidth]{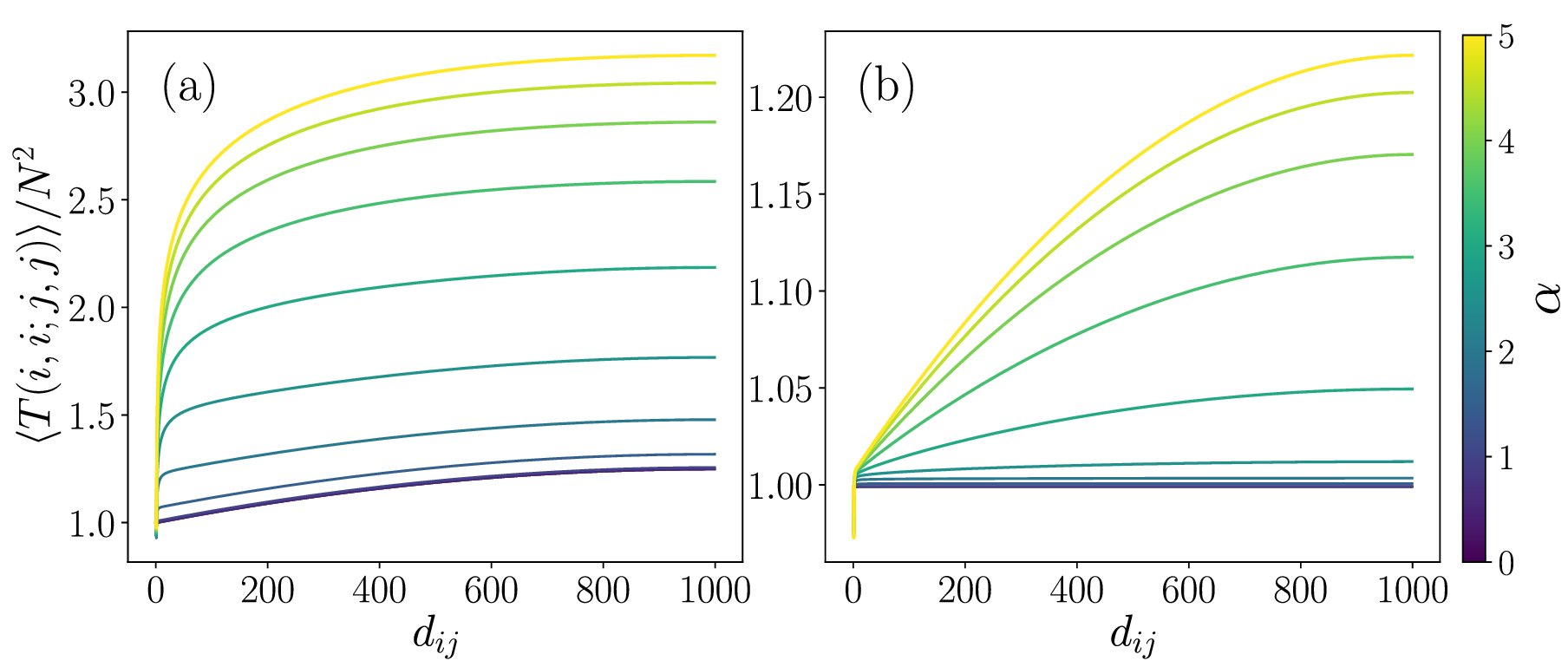}
	\end{center}
	\vspace{-4mm}
	\caption{\label{Fig_3}(Color online) Mean first-encounter times for two synchronous random walkers in a ring with $N=2001$. We show $\langle T(i,i;j,j)\rangle/N^2 $  as a function of $d_{ij}$ for a random walker with L\'evy flights defined by $\alpha=0,0.5,1,\ldots 5$ (in the colorbar) and a second random walker defined by: (a) a normal random walker and (b) L\'evy flights with $\alpha=1$.} 
\end{figure*}
The result in Eq.~(\ref{MFETtworandomkappa1}) is general for the encounter times of two synchronous random walkers when the process is ergodic ($\kappa=1$). To illustrate the variety of possible situations, let us explore the dynamics with L\'evy flights on rings.  L\'evy flights on networks were introduced in Ref.~\cite{RiascosMateos2012} and are the mechanism behind fractional diffusion on networks \cite{RiascosMateosFD2014,FractionalBook2019}. In this case the transition probabilities are defined in terms of the topological distance $d_{ij}$, the number of edges in the shortest path between nodes $i$ and $j$ \cite{RiascosMateos2012}, and are given by
\begin{equation}\label{wijLevy}
w_{i\to j}=\frac{d_{ij}^{-\alpha}}{\sum_{l \neq i}d_{il}^{-\alpha}}\qquad \text{for}\qquad i\neq j 
\end{equation}
and $w_{i\to i}=0$. This random walk allows long-range displacements on the network for  $0\leq\alpha<\infty$; transitions to nearest neighbors have high probability, but hops beyond local nodes are also allowed, generalizing the dynamics observed for normal random walkers. In the limit $\alpha\to\infty$ we have $\lim_{\alpha\to\infty} d_{ij}^{-\alpha}=A_{ij}$, so that $w_{i\to j}=\frac{A_{ij}}{k_i}$ and the L\'evy strategy recovers the normal random walk. When $\alpha\to 0$,  $\lim_{\alpha\to 0} d_{ij}^{-\alpha}=1$ if $i\neq j$ and the dynamics reaches any node with equal probability and equivalent to a normal random walker on a fully connected graph \cite{RiascosMateos2012}. The stationary distribution of a single random walker following L\'evy flights is given by \cite{RiascosMateos2012,riascos2019random}
\begin{equation}\label{PinfLevy}
P_i^{\infty}=\frac{\mathcal{S}_{i}(\alpha)}{\sum_{l=1}^N \mathcal{S}_{l}(\alpha)}\qquad \text{with}\quad \mathcal{S}_{i}(\alpha)\equiv\sum_{m\neq i}d_{im}^{-\alpha}.
\end{equation}
Here $\mathcal{S}_{i}(\alpha)$ is the \emph{long-range degree} that satisfies \cite{RiascosMateos2012}
\begin{equation}\label{Ssum}
\mathcal{S}_{i}(\alpha) =  \sum_{l=1}^{N-1} \frac{1}{l^{\alpha}} k_i^{(l)}=k_i+ \frac{k_i^{(2)}}{2^{\alpha}}+ \frac{k_i^{(3)}}{3^{\alpha}}+\cdots   ,
\end{equation}
where $ k_i^{(n)}$ is the number of $n$-nearest neighbors of the node $i$. The results in Eqs.~(\ref{PinfLevy})--(\ref{Ssum}) show how the stationary probability $P_i^{\infty}$ incorporates information about the network taking into account nodes at different distances from $i$.
\\[2mm]
We proceed to explore encounter times of two walkers following L\'evy flight dynamics on networks. We analyze the dynamics on a ring (finite cycle with periodic boundaries) with $N$ nodes, for which we can deduce analytical expressions for $\langle T(i_1,i_2;j,j)\rangle$. In this case, the long-range degree $\mathcal{S}_i(\alpha)=\mathcal{S}(\alpha)$ is the same for all nodes; as a consequence, the transition matrix $\mathbf{W}$ for each random walk strategy has the structure of a \emph{circulant matrix}, for which all eigenvalues and eigenvectors are well known  \cite{Aldrovandi2001,VanMieghem2011}. In a circulant matrix $\mathbf{C}$ of size $N \times N$ with elements $C_{ij}$, each column has real elements $c_0, c_1,\ldots,c_{N-1}$, ordered in such a way that $ c_0 $ describes the diagonal elements and $C_{ij}=c_{(i-j)\text{mod}\, N}$. In this symmetric matrix, the right eigenvectors $\{|\Psi_m\rangle\}_{m=1}^{N}$ have  components $\langle l |\Psi_m\rangle=\frac{1}{\sqrt{N}}e^{-\mathrm{i}\frac{2\pi}{N}(l-1)(m-1)}$, where $\mathrm{i}=\sqrt{-1}$ (see Ref.
~\cite{VanMieghem2011} for details). These eigenvectors $|\Psi_l\rangle$ satisfy $\mathbf{C}|\Psi_l\rangle=\eta_l |\Psi_l\rangle$, where the eigenvalues $\eta_l$ are given by \cite{VanMieghem2011}
\begin{equation} \label{SpectCeta}
\eta_l=\sum_{m=0}^{N-1} c_m \exp\left[\mathrm{i}\frac{2\pi}{N}(l-1)\, m\right]
\end{equation}
for $l=1,2,\ldots, N$. This result defines the eigenvalues of $\mathbf{C}$ in terms of the coefficients $c_0, c_1,\ldots,c_{N-1}$.
\\[2mm]
On the other hand, for L\'evy flights on rings we have the transition probabilities for $i\neq j$
\begin{equation}
w_{i\to j}=\frac{d_{ij}^{-\alpha}}{\mathcal{S}(\alpha)},
\end{equation}
where distances $d_{ij}$ on the ring satisfy the relation $\cos\left[\frac{2\pi}{N}d_{ij}\right]=\cos\left[\frac{2\pi}{N}(i-j)\right]$ \cite{RiascosMateosFD2015}. Therefore, we can define $\mathbf{W}$ with the coefficients $c_0=0$ and $c_{m-1}=d_{1m}^{-\alpha}/\mathcal{S}(\alpha)$ for $m=2,\ldots,N$. Using this definition and Eq. (\ref{SpectCeta}), we have for the eigenvalues
\begin{equation}
\lambda_l(\alpha)=\frac{1}{\mathcal{S}(\alpha)}\sum_{m=2}^{N}d_{1m}^{-\alpha}\exp\left[\mathrm{i}\frac{2\pi}{N}(l-1)(m-1)\right].
\end{equation}
Having obtained the eigenvalues of the transition matrix $\mathbf{W}$, we analyze the dynamics of two walkers following the L\'evy flight strategy with $0\leq \alpha_1<\infty$ for the first walker and $0\leq \alpha_2<\infty$ for the second one. In this case, $\kappa=1$ and we can apply Eq. (\ref{MFETtworandomkappa1}) to calculate the mean first-encounter times $\langle T(i_1,i_2;j,j)\rangle$. Also, for rings the long-range degree is the same for all the nodes, so that $P_i^\infty=1/N$ and using the eigenvectors of a circulant matrix we obtain
\begin{equation}
X_{ij}^{(l)}=\left\langle i|\Psi_l\right\rangle \left\langle \Psi_l|j\right\rangle=\frac{1}{N}\exp\left[\mathrm{i}\frac{2\pi}{N}(l-1)(j-i)\right].
\end{equation}
Therefore, Eq. (\ref{MFETtworandomkappa1}) for $i_1=i_2=j$ gives 
\begin{equation}
\langle T(j,j;j,j)\rangle=N^{2}.
\end{equation}
In other cases
\begin{multline}
\langle T(i_1,i_2;j,j)\rangle=\sum_{l,m=1}^N g(\lambda_l(\alpha_1)\lambda_m(\alpha_2))\times\\
\left(1-e^{\mathrm{i}\frac{2\pi}{N}(l-1)(j-i_1)}e^{\mathrm{i}\frac{2\pi}{N}(m-1)(j-i_2)}\right).
\end{multline}
Finally, we can apply an additional simplification considering the same initial node for the two walkers, i.e., $i_1=i_2=i$. Hence, for $i\neq j$

\begin{multline}
\langle T(i,i;j,j)\rangle=\\\sum_{l,m=1}^N g(\lambda_l(\alpha_1)\lambda_m(\alpha_2))\left(1-e^{\mathrm{i}\frac{2\pi}{N}(j-i)(l+m-2)}\right).
\end{multline}
In Fig.~\ref{Fig_3}, we show the results obtained for two random walkers $A$ and $B$ on a ring with $N=2001$ nodes. We calculate the average times $\langle T(i,i;j,j)\rangle$ for walkers starting at node $i$ that coincide for the first time at node $j$; these values are presented as a function of the distance $d_{ij}$. In the cases explored, the activity of the random walker $A$ is defined by L\'evy flights with different values of $\alpha$. For the second random walker, in Fig.~\ref{Fig_3}(a) $B$ is taken to be a normal random walker (limit $\alpha\to\infty$); in this case the results show that the best strategy to find the normal random walker is to use L\'evy flights with small values of $\alpha$, for example $\alpha=0,0.5,1$. In Fig.~\ref{Fig_3}(b) $B$ follows L\'evy flights with $\alpha=1$. In this case, as a consequence of the non-locality of the dynamics for $\alpha$ small, for $\alpha=0,0.5$ the encounter times are approximately independent of the distance; however, when we increase $\alpha>2$ the distance between the initial node and the node where the two synchronous walkers coincide becomes relevant, as we also observe in Fig.
~\ref{Fig_3}(a).
\subsection{Dynamics on regular combs}
\begin{figure*}[!t] 
	\vspace{3mm}
	\begin{center}
		\includegraphics*[width=0.9\textwidth]{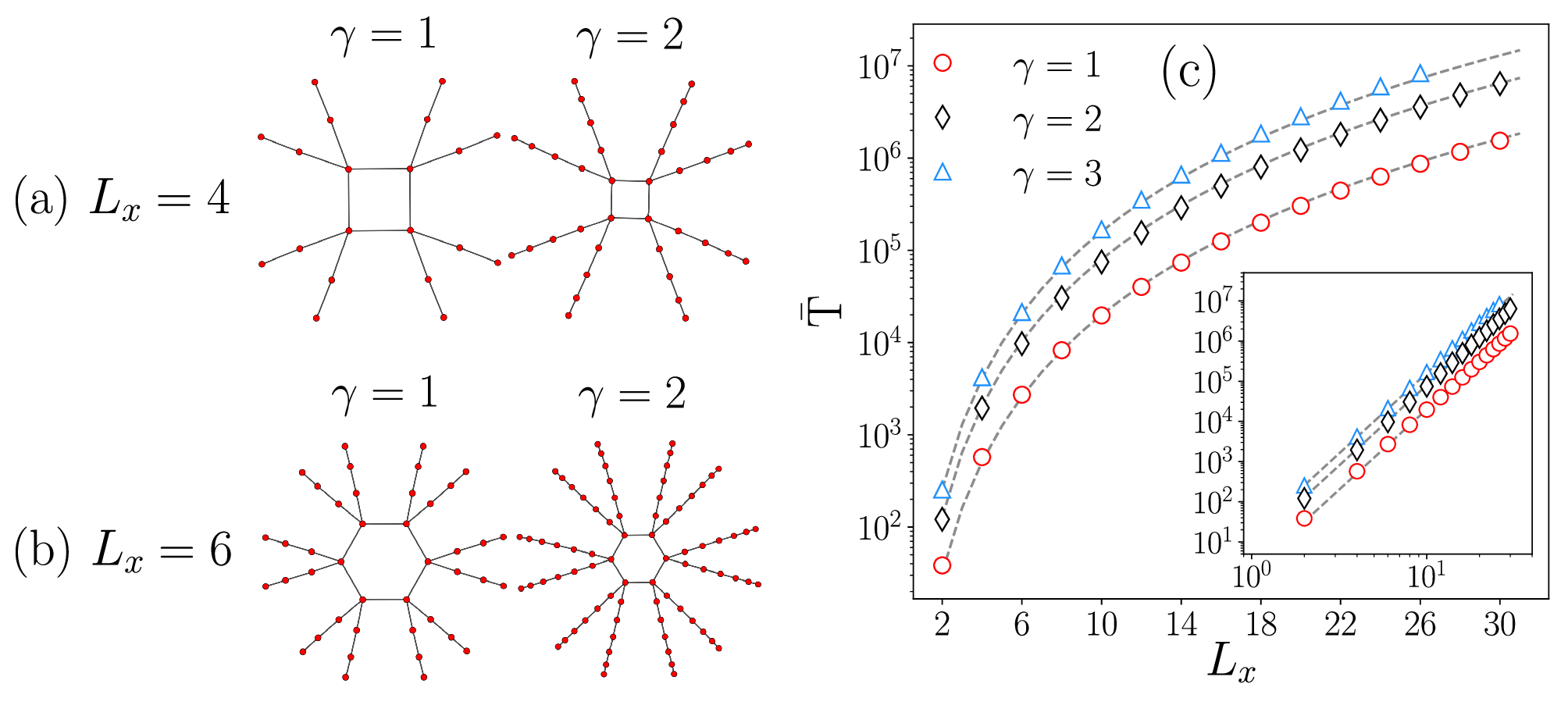}
	\end{center}
	\vspace{-3mm}
	\caption{\label{Fig_4}(Color online) Regular combs with different values of $L_x$ and $\gamma$. Networks with (a) $L_x=4$ and (b) $L_x=6$. (c) Global time $\bar{\mathrm{T}}$ for the average encounter times of two synchronous normal random walkers on regular combs. For each network we calculate $\bar{\mathrm{T}}$ using Eq. (\ref{GlobalT2}) for different values of $L_x$ and $\gamma$ that define each network. Dashed curves represent the relation $\bar{\mathrm{T}}\propto L_x^4$, the inset shows the results in logarithmic scale.} 
\end{figure*}
Having in hand analytical expressions for the mean first-encounter time of two random walkers $\langle T(i_1,i_2;j,j)\rangle$, we explore a \emph{global} time, obtained from the average of these quantities over all  nodes. One alternative is to define a mean time giving the average of $\langle T(i_1;i_2;j,j)\rangle$ over all the possible initial nodes of the two walkers. However, as we mentioned before, there may exist conditions in which the random walkers never coincide in a node, e.g., for two normal walkers on a bipartite network. This motivates the introduction of a time $\mathrm{T}_j$ giving the average of $\langle T(i_1,i_2;j,j)\rangle$  considering that the two random walkers start from the same node, i.e. $i_1=i_2=i$:
\begin{equation}
 \mathrm{T}_j=\frac{1}{N}\sum_{i=1}^N \langle T(i,i;j,j)\rangle.
\end{equation}
$ \mathrm{T}_j$ is an estimate of the number of steps needed  to start at the same node and re-encounter one another at node $j$. In a similar way, we define the global time $\bar{\mathrm{T}}$ by
\begin{equation}\label{GlobalT2}
\bar{\mathrm{T}}=\frac{1}{N}\sum_{j=1}^N \mathrm{T}_j=\frac{1}{N^2}\sum_{i,j=1}^N \langle T(i,i;j,j)\rangle.
\end{equation}
We analyze this global time for regular combs \cite{AgliariPRE2014}, i.e., branched structures obtained from a ring of size $L_x$ (for simplicity chosen even) by attaching to each node two side chains of length $L_y/2$. In addition, the value $L_y$ is defined as $L_y=\gamma L_x$ for $\gamma=1,2,3,\ldots$. The resulting structure is a bipartite graph with $N=L_x(\gamma L_x+1)$ nodes. In Figs.~\ref{Fig_4}(a)-(b) we present some examples of regular combs with $L_x=4,6$ and $\gamma=1,2$. 
\\[2mm]
The study of diffusion and random walkers on combs has been addressed by different authors and recently has been studied in the context of encounter times \cite{AgliariPRE2014,AgliariPRE2019}. Our analytical approach can be used to obtain global times that characterize the synchronous dynamics of two normal random walkers in a regular comb. In Fig.~\ref{Fig_4}(c) we depict the results obtained for the global time $\bar{\mathrm{T}}$ for different values of $L_x$ and $\gamma$, including networks with several sizes, from $N=6$ (for $\gamma=1$ and $L_x=2$) to $N=2054$ (for $\gamma=3$ and $L_x=26$).  We observe how in the range of values explored, the time $\bar{\mathrm{T}}\propto L_x^4$. A similar result was obtained using Monte Carlo simulations from a different approach explored by Agliari et. al. in Ref. \cite{AgliariPRE2014}.
\subsection{Degree biased random walks}
Now, we discuss encounter times of two agents following local degree biased random walks. In this case, a single random walker hops with local transition probabilities $w_{i \to j}$ depending on the degrees of the neighbors of node $i$. Degree biased random walks are defined by \cite{FronczakPRE2009} 
\begin{equation}\label{wijBRW}
w_{i\rightarrow j}=\frac{A_{ij} k_j^{\beta}}{\sum_{l=1}^N A_{il} k_l^{\beta}},
\end{equation}
where $\beta$ is a real parameter. In Eq.~(\ref{wijBRW}), $\beta>0$ describes the bias to hop to neighbor nodes with a higher degree, whereas for $\beta<0$ this behavior is inverted and, the walker tends to hop to nodes less connected. When $\beta=0$, the normal random walk strategy is recovered. In connected undirected networks, degree  biased random walks are ergodic for $\beta$ finite, with stationary distribution
\begin{equation}\label{StatPBRW}
P_i^{\infty}=\frac{\sum_{l=1}^N (k_i k_l)^\beta A_{il}}{\sum_{l,m=1}^N (k_l k_m)^\beta A_{lm}} \, .
\end{equation}
Degree  biased random walks have been studied extensively in the literature in different contexts as varied as routing processes \cite{WangPRE2006}, chemical reactions \cite{KwonPRE2010}, extreme events \cite{KishorePRE2012,LingEPJB2013}, among others \cite{FronczakPRE2009,LambiottePRE2011,Battiston2016}. Additionally, mean field approximations have been explored for diverse cases \cite{FronczakPRE2009,KwonPRE2010,ZhangJSMTE2011}.
\\[2mm]
We analyze the synchronous dynamics of two random walkers on a scale-free network, generated with the preferential attachment algorithm \cite{barabasi2016book,BarabasiAlbert1999}. The random walkers are independent and defined by  Eq. (\ref{wijBRW}) with $\beta=\beta_1$ for the first random walker and $\beta=\beta_2$ for the second one. In a similar way to the cases analyzed before, we define a global encounter time with the average over all the initial positions; however, due to the heterogeneity of the nodes in this network, we weight the values $\langle T(i_1,i_2;j,j)\rangle$ for the initial conditions $i_1$ and $i_2$ with the stationary distributions $P_{i_1}^{1,\infty}$ and  $P_{i_2}^{2,\infty}$ (given analytically by Eq. (\ref{StatPBRW}) with $\beta_1$ and $\beta_2$, respectively). In this way, the average encounter time at node $j$ is
\begin{figure}[!t] 
	\begin{center}
		\includegraphics*[width=0.475\textwidth]{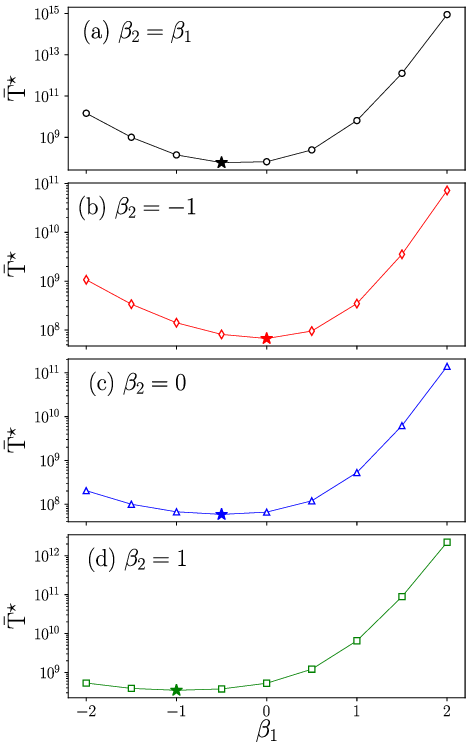}
	\end{center}
	\vspace{-5mm}
	\caption{\label{Fig_Pref}(Color online) Synchronous random walkers on a scale free network with $N=5000$ nodes. We show the average global time $\hat{\mathrm{T}}^\star$ in Eq.~(\ref{GlobalTimeDBRW}) for two degree biased random walkers $r=1,2$ defined by the transition probabilities in Eq.~(\ref{wijBRW}) with $\beta=\beta_1$ and $\beta=\beta_2$, respectively. We present the numerical results for $\beta_1=-2,-1.5,\ldots,1.5,2$ and: (a) $\beta_2=\beta_1$, (b) $\beta_2=-1$, (c) $\beta_2=0$, (d) $\beta_2=1$. Markers represent the numerical values, continuous lines are included as a guide and the minimum values of $\hat{\mathrm{T}}^\star$ for the cases explored are displayed with stars. } 
\end{figure}
\begin{equation}
\mathrm{T}^\star_j \equiv \sum_{i_1=1}^N\sum_{i_2=1}^N P_{i_1}^{1,\infty}P_{i_2}^{2,\infty}\langle T(i_1,i_2;j,j)\rangle.
\end{equation}
Then, applying the result in Eq.~(\ref{TijSpect}), we have for cases with $\kappa=1$
\begin{align}\nonumber
&\mathrm{T}^\star_j=1+\frac{1}{P_{j}^{1,\infty}P_{j}^{2,\infty}}\left[\sum_{l,m=1}^N g(\lambda_l^{(1)}\lambda_m^{(2)})\times\right.\\\nonumber
&\Bigg(\langle j|\phi^{(1)}_l\rangle \langle \bar{\phi}^{(1)}_l|j\rangle\langle j|\phi^{(2)}_m\rangle \langle \bar{\phi}^{(2)}_m|j\rangle-\langle \bar{\phi}^{(1)}_l|j\rangle \langle \bar{\phi}^{(2)}_m|j\rangle\times\\
&\hspace{5mm}\left.\left. \sum_{i_1=1}^N P_{i_1}^{1,\infty}\langle i_1|\phi^{(1)}_l\rangle \sum_{i_2=1}^N P_{i_2}^{2,\infty}\langle i_2|\phi^{(2)}_m\rangle \right)
\right].\label{MeanTBRW}
\end{align}
However, due to the orthogonality between the eigenvectors $\langle \bar{\phi}^{(r)}_1|$ and $|\phi^{(r)}_l\rangle$ for each random walker $r=1,2$, we have $\sum_{i_r=1}^N P_{i_r}^{r,\infty}\langle i_r|\phi^{(r)}_l\rangle=0$ for $l=2,3,\ldots,N$ and $r=1,2$. Therefore, Eq. (\ref{MeanTBRW}) takes the form
\begin{equation}\label{Taverage_centralj}
\mathrm{T}^\star_j=\sum_{i_1=1}^N\sum_{i_2=1}^N P_{i_1}^{1,\infty}P_{i_2}^{2,\infty}\langle T(i_1,i_2;j,j)\rangle=1+\mathcal{T}_j
\end{equation}
with
\begin{equation}\label{CentralityDBRW}
\mathcal{T}_j\equiv\frac{\sum\limits_{l,m=1}^N g(\lambda_l^{(1)}\lambda_m^{(2)})\langle j|\phi^{(1)}_l\rangle \langle \bar{\phi}^{(1)}_l|j\rangle\langle j|\phi^{(2)}_m\rangle \langle \bar{\phi}^{(2)}_m|j\rangle}{P_{j}^{1,\infty}P_{j}^{2,\infty}}.
\end{equation}
In this way, $\mathcal{T}_j$ is a measure of the average time needed to reach simultaneously the node $j$ from randomly chosen nodes on the network and the quantity $\mathcal{T}_j^{-1}$ is a random walk encounter centrality  at node $j$ for the simultaneous dynamics. This is a general form of the random walk centrality of a single random walker introduced in Ref. \cite{NohRieger}, where a centrality $C_j$ combines information of the network and the random walk strategy implemented to visit nodes and gives a high value to nodes easy to reach and small values to nodes for which the random walker takes, in average, many steps to hit the node for the first time starting from any node of the network \cite{NohRieger,RiascosMateos2012}.
\\[2mm]
Hence, the average of $\mathrm{T}^\star_j$ in Eq.~(\ref{Taverage_centralj}), allows to define the global time
\begin{equation}\label{GlobalTimeDBRW}
\bar{\mathrm{T}}^\star\equiv\frac{1}{N}\sum_{j=1}^N\mathrm{T}^\star_j=1+\frac{1}{N}\sum_{j=1}^N\mathcal{T}_j.
\end{equation}
In Fig.~\ref{Fig_Pref} we analyze the global time $\bar{\mathrm{T}}^\star$ in Eq.~(\ref{GlobalTimeDBRW}) for two degree biased random walkers defined by Eq. (\ref{wijBRW}) with parameters $\beta_1$ (first random walker) and $\beta_2$ (second random walker) on a scale-free network with $N=5000$ nodes. We examine different combinations with $\beta_1=-2,-1.5,\ldots,1.5,2$. First, in Fig.~\ref{Fig_Pref}(a) the second random walker is defined by $\beta_2=\beta_1$; in this way, the two random walkers follow the same strategy. We see that, for the cases explored, $\beta_1=\beta_2=-0.5$ minimize the global average encounter time $\bar{\mathrm{T}}^\star$. This result shows that a small bias to visit nodes with lower connections favors the encounters reducing the average first encounter time. In contrast, when $\beta_1=\beta_2=2$ the walkers prefer to hop to nodes with the highest degree and, although this can be a good strategy to reach easily these nodes, at a global scale it is seen that it does not favor fast encounters on the whole network, increasing the value $\bar{\mathrm{T}}^\star$. 
\\[2mm]
In the results in Figs.~\ref{Fig_Pref}(b)-(d), the first random walker is defined with $\beta_1=-2,-1.5,\ldots,1.5,2$ and the second one takes the values $\beta_2=-1,0,1$. The numerical results for each case show that as $\beta_2$ increases, the $\beta_1$ that minimizes $\bar{\mathrm{T}}^\star$ decreases. For example, in Fig.~\ref{Fig_Pref}(b), while $\beta_2=-1$ produces a bias to nodes with fewer connections, the walker that optimizes $\bar{\mathrm{T}}^\star$ occurs when $\beta_1=0$, that is, when there is no bias. In addition, in  Fig. \ref{Fig_Pref}(c), when exploring the cases without bias with $\beta_2=0$, a better result for $\bar{\mathrm{T}}^\star$ occurs when $\beta_1=-0.5$, we see also that $\beta_1=-1$ and $\beta_2 = 0$ have similar times $\bar{\mathrm{T}}^\star$. Finally, in Fig.~\ref{Fig_Pref}(d) with $\beta_2=1$ there is a marked bias of the second random walker towards highly connected nodes. In this case, the smallest values of $\bar{\mathrm{T}}^\star$ are found when $\beta_1\leq0$, the optimal value is obtained for $\beta_1=-1$.
\subsection{$R$ random walkers}
\begin{figure*}[!t] 
\begin{center}
		\includegraphics*[width=1.0\textwidth]{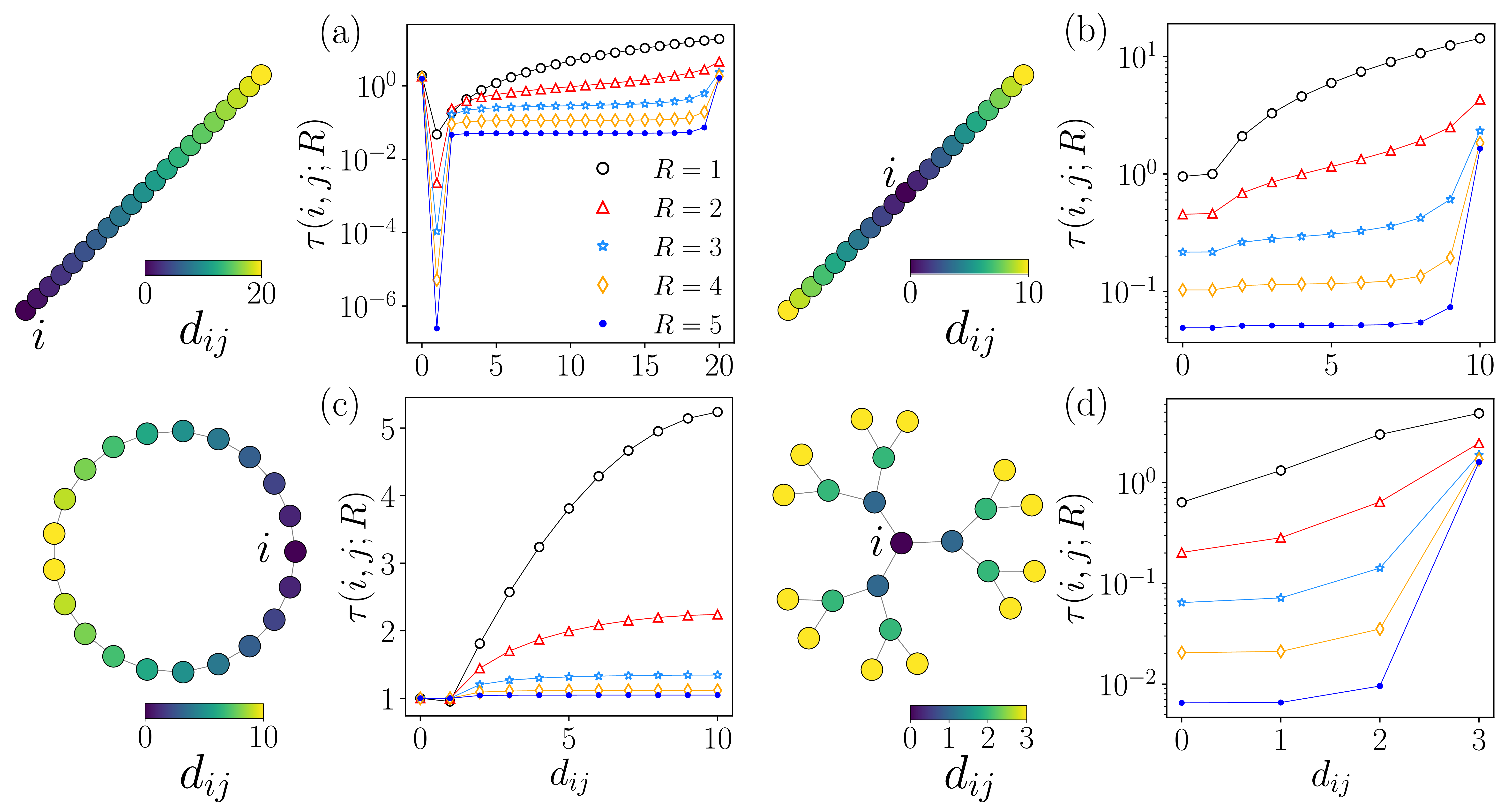}
\end{center}
\vspace{-3mm}
\caption{\label{Fig_6}(Color online) Mean first-encounter times for $R$ synchronous normal random walkers. We show the values $\tau(i,j;R)$ in Eq.~(\ref{TauTimeij}) in terms of the distance $d_{ij}$ for $R=1,2,\ldots,5$ random walkers. Here $\langle T(i\ldots i;j\ldots j)\rangle$ is the average time required for the $R$ random walkers to start in the node $i$ and coincide for the first time in the node $j$, for $R=1$ the $\left\langle T(i; j)\right\rangle$ is the mean first-passage time. (a) Linear graph with initial node $i$ at one of the limits of the network whereas in (b) the initial node $i$ is the central node. In  (c) we explore a ring and (d) a Cayley tree with initial condition $i$ in the central node. In all these cases we present the network and the distance $d_{ij}$ between nodes $i$ and $j$ is represented in the colorbar.} 
\end{figure*}
The result for the average time $\langle T(\vec{i};\vec{j}\,)\rangle$ in Eq.~(\ref{TijSpect}) is general and applies for $R$ non-interacting random walkers in connected networks when each random walker can reach any node of the network from any initial condition. The formalism is also valid for a single random walker, in this case, $R=1$ and the mean first-passage time $\langle T(i;j)\rangle$ expressed in terms of eigenvalues and eigenvectors of the transition matrix $\mathbf{W}$ is recovered. On the other hand, for $R=2,3\ldots$, Eq. (\ref{TijSpect}) gives average times to start at a particular configuration and reach specific nodes for the first time. In the following, we extend our analysis of mean first-encounter times to $R$ synchronous random walkers.
\\[2mm]
To compare the encounter times of $R$ walkers starting at $t=0$ in the node $i$ and meeting for the first time at node $j$, we analyze the scaled time $\tau(i,j;R)$ given by
\begin{equation}\label{TauTimeij}
\tau(i,j;R)=\frac{\left\langle T(i,i,\ldots,i;j,j,\ldots, j)\right\rangle}{N^R}.
\end{equation}
Here, $N$ is the number of nodes in the network and $\left\langle T(i\ldots i;j\ldots j)\right\rangle$ is obtained 
using Eq.~(\ref{TijSpect}).
\\[2mm]
In Fig.~\ref{Fig_6} we analyze $\tau(i,j;R)$ for $R=1,2,\ldots, 5$ synchronous normal random walkers on different network topologies. The results are shown as a function of the distance $d_{ij}$ between the initial node $i$, where all the random walkers start, and the node $j$ where they coincide. In Figs.~\ref{Fig_6}(a)-(b) we have a linear graph with $N=21$ nodes using two initial conditions. In Fig.~\ref{Fig_6}(a) the walkers start at one end of the network; from this node, the agents reach the neighboring node in one step, so that $\tau(i,j;R)=1/N^R$ for $d_{ij}=1$. For $1<d_{ij}<20$ we see how $\tau(i,j;R)$ increases, with a maximum when they coincide at the opposite end of the line. 

In Fig.~\ref{Fig_6}(b) we explore the same linear graph, but now choosing the initial node at the center of the network; the results show that the encounter times differ significantly with the change of the initial condition. 
\\[2mm]
In Fig.~\ref{Fig_6}(c) we analyze a ring with $N=21$ nodes. In this regular structure the stationary distribution for each random walker is $P_i^\infty=1/N$ and the time required to re-encounter in the initial node gives $\tau(i,j;R)=1$; other results for this case can be explored analytically using the approach of circulant matrices presented before for L\'evy flights on rings. 
\\[2mm]
In Fig.~\ref{Fig_6}(d), we have a Cayley tree with $N=22$ nodes and initial node in the center of the tree. Due to the symmetry of the structure, random walkers coincide at the same time in nodes located at the same distance of the center, independently of the branch.
\\[2mm]
In the results in Fig.~\ref{Fig_6} we also observe the effect of the degeneracy $\kappa$ of the highest eigenvalue $\zeta=1$. Since  the linear graph and the Cayley tree are bipartite networks, we have the eigenvalue $\lambda=-1$ for each of the matrices $\mathbf{W}$ defining the normal random walker. In this way $\kappa=2$ for two random walkers, as described previously; in the general case $\kappa=2^{R-1}$. This value modifies the stationary distribution and also has an important effect on the average times in 
Eq.~(\ref{TijSpect}). In the case of the ring with an odd number of nodes, the network is non-bipartite and as a consequence $\kappa=1$. 

\section{Mean first-encounter times for asynchronous motion}
In this section, we discuss mean first-encounter times for $R$ asynchronous random walks defined by a transition matrix $\mathcal{W}^\mathrm{A}$ given by 
Eq.~(\ref{Wmatrix_A}). Recall from
Sec.~\ref{Sec_General} that in the asynchronous setting at each time $t=1,2,\ldots$, one random walker is chosen randomly with equal probability $1/R$ and moves following its particular transition matrix.  Although this motion is completely different from the synchronous dynamics, the analytical result for the mean first-passage times in 
Eq.~(\ref{TijSpect}) has the same form, but now we must use the eigenvalues $\zeta_{\vec{l}}$ given by Eq.~(\ref{zeta_A}). Important consequences are derived from this choice. First, the maximum eigenvalue is $\zeta^\mathrm{A}=1$ is unique, i.e. $\kappa=1$; thus the walkers can meet in any node, independently of the initial condition, defining a global ergodic process. In addition, the stationary distribution is the product of the stationary distributions of each random walker, given by Eq.~(\ref{Pstationary_kappa1}).
\\[2mm]
Due to the definition of the asynchronous dynamics, it is clear that mean first-encounter times will, in general, be longer than the results obtained for the synchronous motion, since in the latter case there is much more activity of the walkers, increasing with the number $R$, in contrast to the asynchronous motion, in which only a single walker moves at each step.  Much of the differences between the two forms of movement will depend on the initial conditions and the types of random walkers. In this way, to quantify the results for
Eq.~(\ref{TijSpect}) for these two types of movements, we introduce the ratio
\begin{equation}\label{chi_frac_ij}
\chi(i,j;R)=\frac{\left\langle T(i,i,\ldots, i;j,j,\ldots, j)\right\rangle_{\mathrm{A}}}{\left\langle T(i,i,\ldots, i;j,j,\ldots, j)\right\rangle_{\mathrm{S}}}\, ,
\end{equation}
where $\left\langle T(i,i,\ldots, i;j,j,\ldots, j)\right\rangle_{\mathrm{A}}$ is the mean first-encounter time for $R$ asynchronous random walkers starting from the node $i$ and meeting for the first time in the node $j$, obtained from Eq.~(\ref{TijSpect}) with eigenvalues from Eq.~(\ref{zeta_A}).
Similarly, $\left\langle T(i,i,\ldots, i;j,j,\ldots, j)\right\rangle_{\mathrm{S}}$ refers to the same quantity but evaluated for the synchronous case by using the eigenvalues from Eq.~(\ref{zeta_S}).
\\[2mm]
\begin{figure}[!t] 
	\begin{center}
		\includegraphics*[width=0.47\textwidth]{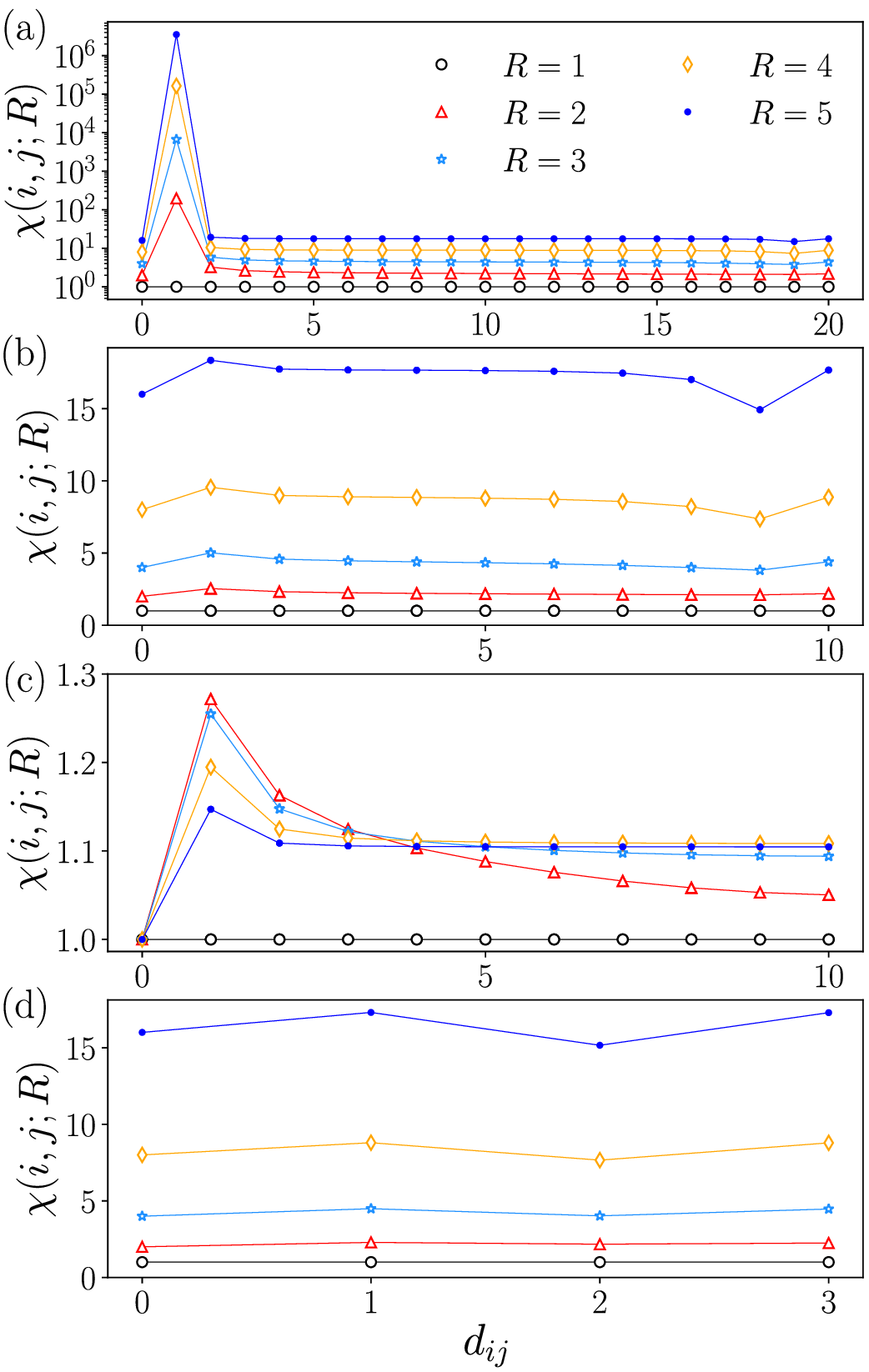}
	\end{center}
	\vspace{-5mm}
	\caption{\label{Fig_7}(Color online) Relation between synchronous and asynchronous motion of $R$ normal random walkers. We calculate the numerical values of $\chi(i,j;R)$ in Eq.~(\ref{chi_frac_ij}) as a function of the distance $d_{ij}$ for $R=1,2,\ldots,5$ random walkers. We explore the four cases analyzed in Fig.~\ref{Fig_6}: (a) a linear graph with initial node $i$ at one of the limits of the network whereas in (b) the initial node $i$ is the central node, (c) a ring and (d) a Cayley tree with initial condition $i$ in the central node. } 
\end{figure}
In Fig.~\ref{Fig_7} we show the values of $\chi(i,j;R)$ for $R=2,\ldots,5$ normal random walkers, analyzing the situations explored in Fig.~\ref{Fig_6}; as a reference we also include the results for $R=1$, giving the horizontal line $\chi(i,j;1)=1$. 
\\[2mm]
In Fig.~\ref{Fig_7}(a) we show the case of the linear graph with initial node $i$ at one of the ends; at this end $\chi(i,i;R)=2^{R-1}$, due to the factor $\kappa=2^{R-1}$ in the stationary distribution for the synchronous motion. On the other hand, for $d_{ij}=1$ in the synchronous dynamics the random walkers always coincide at the first step; however, for the asynchronous motion the result is completely different (for $R=2,3,\ldots,5$), taking a considerable number of steps to coincide in the first neighbor of this end, especially when $R\gg 1$. For $d_{ij}>1$ differences between the times for the asynchronous and synchronous motions are due to the factor $2^{R-1}$, but also depend on the eigenvalue combinations in 
Eqs.~(\ref{zeta_S})--(\ref{zeta_A}). In Fig.~\ref{Fig_7}(b) we analyze the linear graph, but now with the initial condition $i$ at the central node; the main variations in $\chi(i,j;R)$ are associated with the factor $2^{R-1}$, a proportion in which the two stationary distributions differ. In the case of a ring reported in 
Fig.~\ref{Fig_7}(c), we have $\chi(i,i;R)=1$, since the stationary distributions coincide in the synchronous and asynchronous motions; in contrast with the results in Figs.~\ref{Fig_7}(a)-(b), the values $\chi(i,j;R)$ have small variations, maintaining the results close to one. For the Cayley tree with an initial condition in the central node analyzed in Fig.~\ref{Fig_7}(d) we observe a behavior similar to that in Fig.~\ref{Fig_7}(a) for the linear graph.
\section{Encounter times for synchronous motion of bicycles}
\begin{figure*}[!t]
    \begin{center}
        \includegraphics*[width=0.8\textwidth]{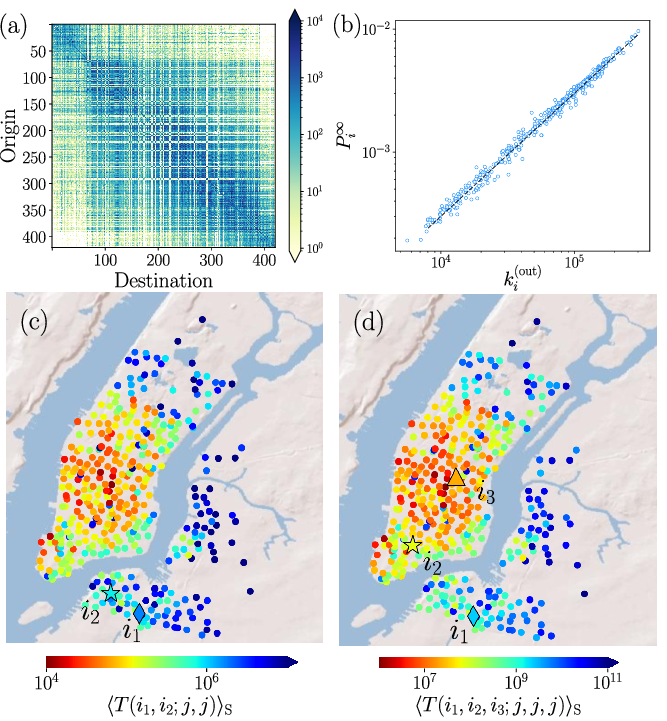}
    \end{center}
    \vspace{-4mm}
    \caption{\label{Fig_8}(Color online) Encounter times in the bike-sharing system Citibike in New York City. Analysis of the activity of $N=421$ stations: (a) Origin-destination matrix with entries $\mathrm{OD}_{ij}$ representing the number of trips from station $i$ ending at $j$, the values are codified in the colorbar, (b) Stationary distribution $P_i^\infty$ of the transition probability matrix defined by Eq. (\ref{wij_OD}) as a function of the out-degree $k_i^{(\mathrm{out})}$, the dashed line represents the relation $P_i^\infty\propto k_i^{(\mathrm{out})}$. Mean-first encounter times for the synchronous dynamics are shown for (c) two (d) three bikes represented with different colors in the map of stations.  Maps were drawn from base maps of satellite imagery (Source: \cite{ArgisMap} and the Matplotlib Basemap package \cite{Basemap}).}
\end{figure*}
In this section, we apply the analytical results for encounter times to the study of synchronous dynamics of bicycles in the bike-sharing system (BSS) Citibike in New York City. The term BSS refers to all the infrastructure and provision of bikes in a system where users pick up and drop off bicycles at self-serving docking stations \cite{Fishman2016}. Each station in the system is represented by a node in a spatial weighted network, where links represent the number of trips between stations. By analyzing data on bicycle trips from June 2013 to December 2016 \cite{CitiBikeDataset}, we obtain an origin--destination matrix, $\mathrm{OD}$, with elements $(\mathrm{OD})_{ij}$ for $i,j=1,\ldots,N$, corresponding to the number of trips starting from station $i$ and ending at $j$. A total of $N=421$ active stations were considered in the analysis of this system; see Ref. \cite{LoaizaMonsalvePlosOne2019} for details on the data processing.

We denote by $k_i^{(\text{out})}$ the total number of bicycles that depart from station $i$ and by $k_i^{(\text{in})}$ the total number arriving at station $i$. In terms of the elements of the $\mathrm{OD}$ matrix we then have
\begin{equation}
k_i^{(\text{out})}=\sum_{\ell=1}^N (\mathrm{OD})_{i\ell},\qquad k_i^{(\text{in})}=\sum_{\ell=1}^N (\mathrm{OD})_{\ell i}.
\end{equation}
The $\mathrm{OD}$ matrix can now be used to define the probability of transition of a bicycle between two stations. Due to the characteristics of this system it is reasonable to approximate it as a Markov process, defined by a stochastic matrix $\mathbf{W}^{(\mathrm{OD})}$ with elements
\begin{equation}\label{wij_OD}
w_{i\to j}^{(\mathrm{OD})}=\frac{(\mathrm{OD})_{ij}}{k_i^{(\mathrm{out})}}.
\end{equation}
The analysis of the Citibike system in Ref.~\cite{LoaizaMonsalvePlosOne2019} reveals a particular relation between the probability $w_{i\to j}^{(\mathrm{OD})}$ and the geographical distance $l_{ij}$ between stations $i$ and $j$. The dynamics described by the transition matrix classify trips as local and long-range transitions. In local displacements, the users travel to stations around a distance $L\approx 1\mathrm{ km}$  from the departure station. In this case, the probability of moving to one of the stations in the local neighborhood is approximately constant. On the other hand, long-range transitions appear for users with displacements to stations beyond the local neighborhood, for which the transition probabilities decay with distance as $w_{i\to j}^{(\mathrm{OD})}\propto l_{ij}^{-2}$, in the same way as in the gravity-law model for human mobility \cite{LoaizaMonsalvePlosOne2019}.
\\[2mm]
For the $N=421$ active stations considered, the dynamical process is ergodic and all the formalism described before for simultaneous random walks on networks can be applied. However, in this case, the eigenvalues of the transition matrix are complex, since the $\mathrm{OD}$ matrix is, in general, not symmetric. The difference between the values $(\mathrm{OD})_{ij}$ and $(\mathrm{OD})_{ji}$ are associated with the accumulation of bikes in particular stations, requiring the massive relocation of bikes between some stations to maintain the correct operation of the whole system, a phenomenon known as re-balancing \cite{MedarddeChardon2016}.
\\[2mm]
In Fig.~\ref{Fig_8} we present our results for the Citibike system. In 
Fig.~\ref{Fig_8}(a) we show the matrix $\mathrm{OD}$, with entries codified as per the color bar; we use this information to define a random walk dynamics with transition probabilities $w_{i\to j}^{(\mathrm{OD})}$ given by 
Eq.~\eqref{wij_OD}. The two eigenvalues of $\mathbf{W}^{(\mathrm{OD})}$ with the largest real part are $\lambda_1=1$ and $\lambda_2=0.7896$. In Fig.~\ref{Fig_8}(b) we show the stationary distribution $P_i^\infty$ of a single random walker, also called \emph{OD-rank} \cite{RiascosMateosSciRep2020}; this gives the importance of a station in the system. The results $P_i^\infty$ are obtained numerically from the left eigenvectors of $\mathbf{W}^{(\mathrm{OD})}$  associated to the eigenvalue $\lambda_1$;  in this way $P_i^\infty\propto \langle\bar{\phi}_1|i\rangle$. Also, we have $P_i^\infty>0$ for all $i$, so that the random walk dynamics is capable of reaching all stations in the system. We represent the stationary distribution in terms of the out-degree $k_i^{(\mathrm{out})}$; the results show that $P_i^\infty\approx \frac{k_i^{(\mathrm{out})}}{\sum_{m=1}^N k_m^{(\mathrm{out})}}$.
\\[2mm]
Now, with the information of the eigenvalues and eigenvectors of the transition matrix $\mathbf{W}^{(\mathrm{OD})}$, we analyze the synchronous dynamics of random walkers. The results from this Markovian approach are a proxy of the real activity of the system that allows the identification of stations with potential accumulation of bikes. In Fig.~\ref{Fig_8}(c) we present the numerical values obtained from Eq.~(\ref{TijSpect}) for the mean encounter times at station $j$ of two synchronous bikes starting from stations $i_1$ and $i_2$ at the south region of New York City; we color each station $j$ with the values  $\langle T(i_1,i_2;j,j)\rangle_\mathrm{S}$, codified in the color bar. 
In Fig.~\ref{Fig_8}(d) we repeat the analysis for three bikes starting from stations  $i_1$, $i_2$,  $i_3$ to obtain $\langle T(i_1,i_2,i_3;j,j,j)\rangle_\mathrm{S}$ for $j=1,2,\ldots,421$. The results show that bicycles will meet faster at stations in the Manhattan zone, where we observe the shortest encounter times. Thus, according to our analysis, stations with the lowest MFET would require more rebalancing.
\\[2mm]
In addition, the analysis of different initial conditions shows that the meeting times of $R$ random walkers are approximated by $\langle T(\vec{i};\vec{j}\,)\rangle\propto \frac{1}{(P_j^\infty)^R}$; the effect of the initial conditions is to introduce small variations to this relation. This result is a consequence of the gap between $\lambda_1$ and $\lambda_2$ that reduces the contribution of the initial conditions in Eq.~(\ref{TijSpect}). In other cases analyzed previously for local random walks on networks with the large-world property (rings and trees), this gap is small, assigning major importance to the initial conditions. 
\\[2mm]
Although our analysis of BSS is an approximation assuming a Markovian dynamics, the results provide a first insight into the collective dynamics in shared bicycle systems.
\section{Conclusions}
In conclusion, we deduced analytical expressions for the study of the dynamics of $R$ non-interacting random walks on networks. Our formalism explores analytically the global dynamics of synchronous and asynchronous motion in terms of the spectral representation of the transition matrices that define independent Markovian random walkers. We illustrate the general results by calculating mean first-encounter times of two synchronous random walkers on different types of networks. For the synchronous motion, we explore normal random walks on a Cayley tree and a Watts--Strogatz random network. Also, we deduce analytical expressions for L\'evy flights on rings, to explore mean first-encounter times for random walkers following different types of random hopping between nodes and global times for two walkers on regular combs. We then analyze encounter times for $R=1,2,\ldots,5$ normal random walkers in a linear graph, a ring, and a Cayley tree and the relation between synchronous and asynchronous dynamics. 
\\[2mm]
We applied our methodology to study the activity of the bike-sharing system Citibike in New York City, where we explore encounter times of two and three bikes at each station. This example shows how the methods introduced are general, and extensions of this work will be useful for applications to human mobility, encounter networks, epidemic spreading, and ecology, among many other fields.
\\[2mm]
This mathematical framework can be applied to other contexts in human mobility, such as the movement of taxis \cite{RiascosMateosSciRep2020}, or temporal networks generated from encounters at points of interest in cities \cite{RiascosMateosPlosOne2017}. A more detailed treatment of these problems requires extending the formalism discussed in this work to the case of continuous-time random walkers.

\section*{Acknowledgments}
A.P.R. acknowledges support from PAPIIT-UNAM grant No.~IN116220.
\onecolumngrid
\providecommand{\noopsort}[1]{}\providecommand{\singleletter}[1]{#1}%

\end{document}